\documentclass{aa}
\usepackage{epsfig}
\usepackage{graphicx}
\begin{document}

\title{Fluctuations of K-band galaxy counts}

\author{M. L\'opez-Corredoira\inst{1}, J. E. Betancort-Rijo\inst{2,3}}
\institute{$^1$ Astronomisches Institut der Universit\"at Basel,
Venusstrasse 7, CH-4102 Binningen, Switzerland\\
$^2$ Instituto de Astrof\'\i sica de Canarias, C/.V\'\i a L\'actea, s/n,
E-38200 La Laguna (Tenerife), Spain\\
$^3$ Departamento de Astrof\'\i sica, Universidad de La Laguna, 
Tenerife, Spain}

\offprints{martinlc@astro.unibas.ch}

\date{Received xxxx / Accepted xxxx}

\abstract{We measure the variance in the
distribution of off-plane ($|b|>20^\circ $)
galaxies with $m_K<13.5$ from the 2MASS K-band survey 
in circles of diameter between 0.344$^\circ $ and 57.2$^\circ $.
The use of a near-infrared survey makes negligible the contribution of
Galactic extinction to these fluctuations.
We calculate these variances within
the standard $\Lambda $CDM model assuming that the sources are
distributed like halos of the corresponding mass,
and it reproduces qualitatively the galaxy counts variance. 
Therefore, we test that the counts can be basically
explained in terms only of the large
scale structure. A second result of this paper is a new method to determine
the two point correlation function obtained by forcing
agreement between model and data. This method does not 
need the knowledge of the two-point angular
correlation function, allows an estimation of the errors (which are low
with this method), and can be used even with incomplete surveys.

Using this method 
we get $\xi (z=0, r<10\ h^{-1}{\rm Mpc})=(29.8\pm 0.3)
(r/h^{-1}{\rm Mpc})^{-1.79\pm 0.02}$,
which is the first measure of the amplitude of $\xi $ 
in the local Universe for the K-band.
It is more or less in agreement with those 
obtained through red optical filters selected samples, but it is 
larger than the amplitude obtained for blue optical filters selected samples. 

}

\authorrunning{L\'opez-Corredoira \& Betancort-Rijo}

\titlerunning{galaxy counts fluctuations}

\maketitle

\section{Introduction}

The distribution of the number of galaxies or clusters of galaxies
in a certain volume $V$ can be used to test the large-scale structure. 
The calculation of probabilities of different structures in a clustering model 
are sometimes amenable to analytical techniques: for instance, for voids
(Betancort-Rijo 1990), density of Abel clusters (Betancort-Rijo 1995), 
number of galaxies in a randomly placed sphere and 
cluster density profiles (Betancort-Rijo \& L\'opez-Corredoira 1996), etc.
The calculation of variance of
the number of galaxies within the cones corresponding 
to the circles in the sky, or another 2-D figure,  
(i.e., galaxy counts) is another possibility,
and even easier to compare with the observations since we do not
need any information about the redshifts of the individual galaxies.
This is precisely the purpose of this paper: the calculation from a model
of the variance of the number of galaxies 
in the different regions of the sky, and to
see how well they fit real data obtained directly from a sky survey.
This will allow us to determine some parameters of the two-point 
correlation function independent of any other method.

One disadvantage of the galaxy counts is that they are affected by Galactic
extinction, and it is difficult to separate the fluctuations due to this
extinction from the real fluctuation in the galaxies distribution.
However, the arrival of near-infrared galaxy surveys such as DeNIS or
2MASS provides us a way to avoid this problem, since the extinction
in K-band, the one to be used in this paper, is 10 times lower than in
V-band. In V-band, the dispersion of
galaxy counts due only to the fluctuations of extinction in scales
around 1 degree is around 30\%
[see eq. (\ref{sigmaext}) and the parameters given in the paragraph
before eq. (\ref{sigmaext})] at galactic latitudes
$|b|>20^\circ$, while in K-band is only $\sim 3$\%. 
Since the fluctuations we measure due to the large-scale structure 
are 25-80\% in the explored scales, it is clear that the 
calculation of the structure 
parameters are very sensitive to the exact knowledge of the galactic extinction
in visible bands; however, in near-infrared, the calculation of the parameters is 
much less sensitive to the extinction and we can be sure that the 
measured fluctuations correspond to the real distribution of galaxies.
Moreover, in the K-band
we have an important advantage: the K-correction is nearly independent
of galaxy type (Mannucci et al. 2001), and this will allow us to make 
the K-corrections without knowing the galaxy type of the sources in 
the survey.

Maps of galaxy counts have been available for quite a long time and 
have been used for several purposes. 
Zwicky (1957, p. 84), for instance, produced
them from the available catalogues in visible bands, 
with a smooth correction of galactic extinction as a function of $b$,
in order to explore the intergalactic extinction.
Rather than obtaining the 
Galactic or intergalactic extinction,
the purpose of this paper is to show that the count variances are
in qualitative good agreement with those obtained assuming that the source
correlation function is equal to that obtained in the standard $\Lambda $CDM
model.
Cumulative galaxy counts (in optical) versus the limiting magnitude were used 
to constrain the departure from homogeneity---fractal distribution---at 
large scales (Sandage et al. 1972). A $\log N(m)\propto 0.6 m$ points out 
total homogeneity and this is more or less observed in near infrared counts 
too: with 2MASS data for low redshift (Schneider et al. 1998, fig. 5) or
other surveys (e.x., K\"ummel \& Wagner 2000). 
Here, we will not explore further the homogeneity at very large scales but the 
clustering at small scales for the local Universe ($z<0.4$). 
The use of galaxy counts to derive the two-point angular
correlation function was already used by Porciani \& Giavalisco (2002); however,
we will recover directly (under some assumptions) the two-point correlation
function of the sources, and, therefore, the relative biasing with respect
to the mentioned halos. We may accurately determine it, 
by considering a slightly modified power law (the 
correlation function is a power law in the most relevant region) and
determining the value of the amplitude and the exponent which lead to
the best agreement.
 
The paper is divided as follows: \S \ref{.galcounts} describes
the observational data and the way to measure the variance in the
distribution of galaxies; \S \ref{.model} explains how
to calculate this variance within a model for the large scale structure;
and \S \ref{.comparison} makes the comparison between data and model
predictions and derive the parameters of the
two-point correlation function necessary to get an agreement between model
and data.

\section{Galaxy counts}
\label{.galcounts}

The data used in this work have been taken from the extended sources
of the 2MASS-project (Jarrett et al. 2000), All-sky release
(http://www.ipac.caltech.edu/2mass/releases/docs.html).
Completeness limit: $m_K=13.5$ (Schneider et al. 1998; Jarrett et al. 2000;
Maller et al. 2003, \S 2).
Assuming an average color of $B-K\approx 4$, it is equivalent to an
optical limit of 17.5 (Schneider et al. 1998), deep enough for 
statistical studies of the large scale structure, the local structures
are not too predominant in it, although 
high redshift are excluded [the galaxies have redshifts $z<$0.3-0.4
(Cole et al. 2002), and an average $\langle z\rangle \approx 0.083$ (see 
below for details)].
We do not analyze samples of galaxies below this limit of $m_K$
(for instance, $m_K<12.0$ or $m_K<11.0$, etc) because this would represent
the very local Universe rather than the large scale structure.

\begin{figure*}
\begin{center}
\mbox{\epsfig{file=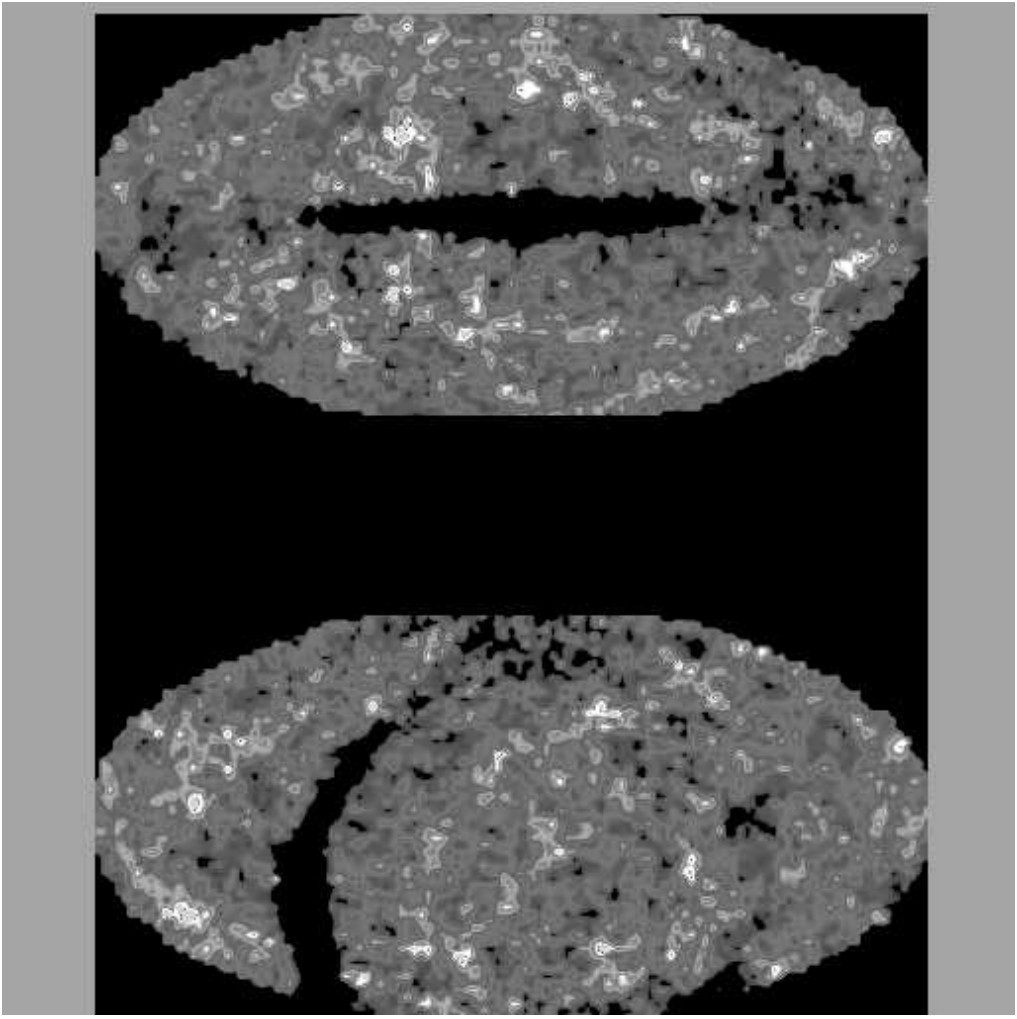,height=14cm}}
\end{center}
\caption{Aitoff projection in galactic coordinates (up) and equatorial
coordinates (down) of the 2MASS/``All-sky release'' 
galaxy counts with $m_K<13.5$; average: 14.4 galaxies/deg$^2$.
Galaxies were counted in square regions of 3 deg.$\times 3$ deg. A 
smoothing of the map was carried out. Black color stands for regions 
with very few galaxies, and high density is represented by white color. 
Extinction effects are negligible in the fluctuations. 
Poissonian noise is not negligible but it is relatively small compared to 
the intrinsic fluctuations due the large scale-structure (see text).}
\label{Fig:aitoff}
\end{figure*}

In Fig. \ref{Fig:aitoff}, we see a representation of the fluctuations
at scale of $3^\circ $ in the whole sky (average
around 150 galaxies per area with complete coverage).
In the zone of avoidance, $|b|<20^\circ $, there is a clear defect
of galaxies due to the extinction, small compared to the optical but
not negligible in near plane regions.
The reliability is larger than 99\% in $|b|>20^\circ $ (Schneider et al. 1998).

\begin{figure}
\begin{center}
\mbox{\epsfig{file=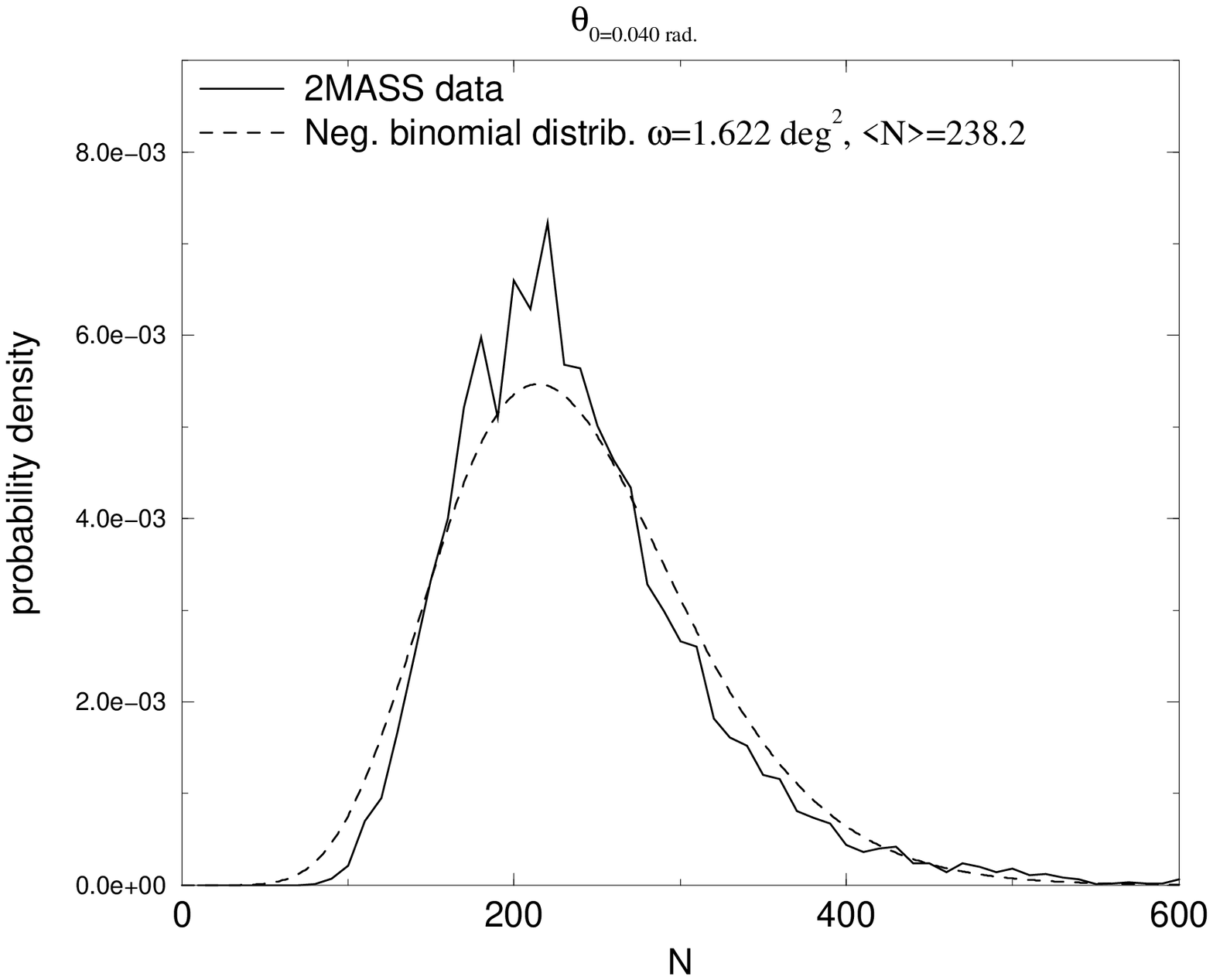,height=6.7cm}}
\end{center}
\caption{Distribution of frequencies of cells with $N$ galaxies for circular
cells of angular radius $\theta _0=0.040$ rad. Binning step: 10.
For comparison, we plot a negative binomial distribution (Betancort-Rijo 2000)
with the same average and variance than the data.}
\label{Fig:histo}
\end{figure}

In order to quantify the fluctuations, we count the number of galaxies
in each circle of sky with angular radius $r_0$. In 3D space, 
it counts the galaxies within the corresponding cone in the line of sight.
We select only the circular regions in $|b|>20^\circ $
which were covered in more than 90\%. The cumulative counts up to magnitude
13.5 are expressed per unit area (we divide the number of galaxies
per region by the area, $S$ of the region).
We measure the counts in randomly placed circular areas instead 
of a regular mesh. Since the number of random circles in which we 
measure the galaxy counts is around 8 times larger than the total area divided 
by the area of the circle, the lost of information is 
very low (a 0.033\% of the galaxies would not be in any circle if the 
distribution were Poissonian).
Once we have these counts for each of the $n$ regions containing
respectively $n_i$ galaxies, we calculate the average, 
\begin{equation}
<N>=\frac{1}{Sn}\left(\sum _{i=1}^{n}n_i\right)\approx 14.4 \ {\rm deg}^{-2} 
,\end{equation}
and the dispersion with respect to the
average, 
\begin{equation}
\sigma _N(\theta _0)=\sqrt{\frac{1}{S^2n}\left(\sum _{i=1}^{n}n_i^2\right)
-\frac{1}{S^2n^2}\left(\sum _{i=1}^{n}n_i\right)^2}
.\end{equation}
This last number gives us information about 
the amount of structure in the 3D cones with angular radius $\theta _0$.
An example of the distribution of counts is given in Fig. \ref{Fig:histo}
and compared with a negative binomial distribution with the same
variance, to which a wide variety of clustering processes 
lead (Betancort-Rijo 2000).

The fluctuations due to the intrinsic clustering of galaxies, 
$\sigma _{\rm st}$,
will be the total fluctuations minus the other independent
sources of fluctuation subtracted quadratically:

\begin{equation}
(\delta N)_{\rm st}=\sqrt{\sigma _N^2-
\sigma _{\rm ext}^2-\sigma _{\rm Poisson}^2}
\label{estrdata}
.\end{equation}

The galactic extinction r.m.s. fluctuations, 
$\sigma _{\rm ext}$, may be evaluated using 
the Schlegel et al. (1998) maps of extinction: in $|b|>20^\circ $, 
the mean extinction in K-band is $A_K=0.019$ mag, and the fluctuations 
of this extinction are $\sigma _{A_K}\approx 0.020$ mag 
(in the scales with $\theta _0$ around 1 degree; it changes slightly
with the scale).
Given that $\log N=constant+0.6m_K$ (for a homogeneous distribution;
see, for instance, Sandage et al. 1972):

\begin{equation}
\sigma _{\rm ext}\equiv \langle (\delta N)^2\rangle _{\rm ext}^{1/2}
=0.6(\ln 10) \sigma _{A_K}<N>
\label{sigmaext}
.\end{equation}
For the K-band $\sigma _N>>\sigma _{\rm ext}$, so the small
uncertainties in the knowledge of the statistical properties
of the extinction are not important.
Note however that in visible bands $\sigma _{\rm ext}$ will be 10 times larger,
comparable to the dispersion due to the large scale structure and, 
therefore, it is
not possible to obtain accurate information about this last dispersion without
an accurate measure of the extinction fluctuations. This is precisely
the advantage of infrared surveys.

The Poissonian fluctuations is the most important contribution, apart
from those due to large scale structure, and it can be exactly determined:
\begin{equation}
\sigma _{\rm Poisson}\equiv \langle (\delta N)^2\rangle _{\rm Poisson}^{1/2}
=\sqrt{<N><1/A>}
.\end{equation}
 
Once the fluctuations due to the structure are obtained by means of eq.
(\ref{estrdata}), these can be compared with the predictions of a 
standard model, as it will be carried out in the following sections.

\section{Model}
\label{.model}

The calculation of the fluctuations $\frac{\delta N}{N}\equiv
\frac{\sqrt{(\delta N)^2_{\rm st}}}{<N>}$ due to the
clustering of the galaxies is carried out in the following way:

\begin{equation}
\langle N\rangle=A \overline{n}\int _{\theta<\theta_0} r^2 \Phi 
[M_K<M_{K,limit}(r)]dr,\end{equation}
\[
(\delta N)^2_{\rm st}=\overline{n}^2
\int _{\theta_1<\theta_0} d\vec{r}_1\int _{\theta_2<\theta_0}\
\Phi [M_K<M_{K,limit}(r_1)] 
\]\begin{equation}\times
\Phi [M_K<M_{K,limit}(r_2)]
\xi [|\vec{r}_1-\vec{r}_2|,z(r_1)]d\vec{r}_2
,\end{equation}
where $\vec{r}$ stands for the radial distance 
in comoving coordinates,
$A =2\pi (1-\cos \theta _0)$ is the area (in steradians)
of the circular regions, 
$\overline{n}$ is the mean space density of galaxies (which is irrelevant
in this context since it cancels when we calculate $\delta N/N$),
$\Phi [M_K<M_{K,limit}(r)]$ is the cumulative normalized luminosity function
up to absolute magnitude $M_{K,limit}[r_{phys}=r/(1+z)]=m_{K,limit}+5-
5\log r_{phys}-CorrK(r_{phys})$.
The K-correction is $CorrK[r_{phys}=r/(1+z)]=-2.955z(r_{phys})+
3.321z(r_{phys})^2$ for $z<0.40$
(beyond $z=0.40$ we do not see practically any galaxy), 
for any galaxy type (obtained from the fit of data from Mannucci et al. 
2001). The redshift is $z(r_{phys})$ corresponds to the
velocity $v(r_{phys})=100r_{phys}(h^{-1}{\rm Mpc})$ km/s
(linear Hubble law taken as an approximation for low redshift galaxies).
The luminosity function $\Phi $ is taken from Kochanek et al. (2001).
We neglect the evolution of the luminosity function, which is
mild ($\frac{dM_K}{dz}\approx 0.5$, Pozzetti et al. 2003)
and very small at the mean redshift of our galaxies ($\langle z
\rangle \approx 0.083$): a shift of $\sim 0.04$ magnitudes
will produce a variation of $\sim 2$\% in both $\langle \delta N\rangle$ 
and $\langle N\rangle $ in the same direction; since we 
calculate the rate $\frac{\delta N}{N}$, this small variations 
mostly compensate, and only second order variations remain, 
which are negligible.
The errors associated with the uncertainty in the
luminosity itself do mainly affect the normalization,
which is also irrelevant for the calculation of $\delta N/N$ since 
normalization factor cancels out.
Finally, the two-point correlation function is taken to be (for comoving
coordinates):

\begin{equation}
\xi (r,z)=\xi(r,z=0)(1+z)^{-(3+\epsilon+\gamma)}
\label{xiev}
,\end{equation}
\begin{equation}
\xi (r,z=0)=
\left \{ \begin{array}{ll} 
	-1,& \mbox{ $r<0.01$} \\ 
	\left(\frac{r}{r_0}\right)^\gamma ,& \mbox{$0.01<r<10$} \\ 
 
	FT[P(K)] ,& \mbox{$r\ge 10$} 
\end{array} 
\right \} 
\label{xiz0}
,\end{equation} 
where $FT[...]$ stands for Fourier transform, and $r$ is given in units
$h^{-1}$Mpc.
The truncation of the two-point correlation function is set at 10 $h^{-1}$kpc,
but the exact value does not matter, our results do not depend on
this minimum scale; it is introduced to avoid computer
algorithm problems at very small scales.
More exactly, the change of regime should occur at the matching
point but, for the interesting range of parameters, this takes place
around 10 $h^{-1}$Mpc. The power spectrum for the linear part 
($r>\approx 10$ $h^{-1}$Mpc), $P(K)$,
is given for a CDM scenario with initial Harrinson-Zeldovich 
power spectra (Bardeen et al. 1986):

\[
P(K)=\frac{A\ln (1+2.34q)^2}{K\sqrt{1+3.89q+(16.1q)^2+(5.46q)^3+
(6.71q)^4}}
,\]\begin{equation}
q=\frac{K}{\Omega _{\rm mat}h\ (h/{\rm Mpc})}
,\end{equation}
and we adopt $A=1.984\times 10^4$ ($\sigma _8=1$) and 
$\Omega _{\rm mat}h=0.21$. Within small scales ($\theta _0\le 0.040$ rad), 
the non-linear part is predominant, so slight variations of the parameter 
$\Omega _{\rm mat}h$
or the normalization $A$ will not lead to changes in our results. 
For instance, numerical experiments have shown us that a variation of 
10\% in the amplitude $A$ leads to variations of $\approx 2$\% in 
$\delta N/N (\theta_0=0.040$ rad) or $\approx 0.7\%$ in
$\delta N/N (\theta_0=0.010$ rad) and much less for lower $\theta _0$;
since the error bars of $\delta N/N $ are of this order, we consider
negligible these small variations due to the change of the parameters
in $P(K)$.
In scales larger than $\theta _0=0.040$ rad, the errors bars are too large
compared to the possible variations in $P(K)$. In all ranges,
even variations up to 10-20\% in $P(K)$ will not affect 
$\delta N/N$ too much compared to the errors.

Two questions might rise as to the suitability of eq. (\ref{xiz0}):
1) is the assumption of a power-law for non-linear scales appropriate?;
2) could we apply a power-law for all scales instead of a $\Lambda $CDM model
for the non-linear regime? Both questions are answered in other papers but
they will also be answered by the result of the fit of the 
counts itself, shown in \S \ref{.comparison} (see Fig. \ref{Fig:fluct_area}).
As to the first question, the fit of the power law with 
$r_0=6.66$ $h^{-1}$Mpc, $\gamma=1.79$
is remarkably good for low $\theta _0$ (which is nearly independent of the 
linear part of $\xi $), a power law in the non-linear
regime gives a very good fit.
The answer to the second question is provided by the following consideration:
a power-law in all scales gives more structure at $r>40$ $h^{-1}$Mpc 
than the power-law in the non-linear regime + $\Lambda $CDM 
model in the linear regime. The first option gives further
fluctuations at large $\theta _0$; the difference is not high enough
to reject the first option (we have already said that the fit is not
very sensitive to the parameters in the linear power spectra), 
but the $\Lambda $CDM model in the linear regime (solid
line in Fig. \ref{Fig:fluct_area}) is considerably better 
than the power-law at all scales (long dashed line 
in Fig. \ref{Fig:fluct_area}). 

There are also theoretical reasons to use expression (\ref{xiz0})
for $\xi (r)$. $\Lambda $CDM simulations lead to a correlation function for
virialized halos with circular velocity $>120 $ km/s with this form and
$\gamma =1.7$ (Primack 2001, Fig. 1). This function turns out to be almost
exactly equal to that for APM galaxies. For K-selected galaxies the correlations
do not need to be equal to that for halos; there may exist some biasing.
However, the biasing ($b(r)=\left(\frac{\xi _K(r)}{\xi
_{halo}(r)}\right)^{1/2}$) is expected to be a mild function of $r$.
So, it seems plausible to use expression (\ref{xiz0}) with different amplitude
and slightly different $\gamma $ with respect to the halos (or APM galaxies).

With this in mind, we shall see in next section that with the assumed
shape in eq. (\ref{xiz0}), there is qualitative agreement with the
observed variances. We consider this result as a confirmation of the
assumption that the measured variances are due to the large scale
structure. This, along with the previously given arguments
supporting this assumption, make us confident as to its correctness.
So, we consider that large scale structure should explain not only the
main part of the variances, but the whole of them. Forcing this by choosing
the appropriate values for $r_0$, $\gamma $ lead us to an alternative
method for determining the correlation function.

The evolution of the correlation function depends on $\Omega _{\rm mat}$ 
and $\Omega _{\rm \Lambda}$ through $\epsilon $. 
In our case, since the average redshift $\langle z\rangle=0.083$ 
(for $m_K<13.5$ using the aforementioned luminosity function and
K-correction), the exact value of $\epsilon $ is not so important, and
small variations on it will not affect significantly the results from the
model. The evolutionary corrections in this small range of redshifts
are also negligible, specially in K-band (Carlberg et al. 1997).
We take the value $\epsilon=-0.1$, which comes from the approximation
of $\xi \propto D(z)^2$, for comoving coordinates
[which holds provided that the shape of $\xi $ does not evolve, i.e. 
$\gamma $ is constant with respect to $z$
(proved by Carlberg et al. 1997)],  
where $D$, the growing factor of the linear density fluctuations, is 
(Heath 1977, Carroll et al. 1992) given by
\begin{equation}
D(z)=\frac{5\Omega _{\rm mat}(t_0)(1+z)}{2f[R(z)]}
\int _0^{R}f^3(R)dR,
\end{equation}\[
R=\frac{1}{1+z}
\]\[
f(R)=\left[1+\Omega _{\rm mat}(t_0)\left(\frac{1}{R}-1\right)
+\Omega _{\rm \Lambda}(R^2-1)\right]^{-1/2}
.\] 
This leads to a value of $\epsilon\approx -0.1$ for 
$\Omega _{\rm mat}(t_0)=0.3$, $\Omega _{\rm \Lambda}=0.7$.
Although, as we have said before,
this value changes with other cosmological parameters, 
our outcome will be practically independent of these small changes.
$\epsilon $ was measured by Carlberg et al. (1997)
using a K-band deep survey, and although the accuracy was not very high,
they obtained a value ($\epsilon=0.2\pm 0.5$) which is compatible 
with $\epsilon =-0.1$. Also, Roche et al. (1999) obtained $\epsilon\approx
0$.

\section{Comparison between model and data. Fit of $r_0$ and $\gamma $}
\label{.comparison}

In the previous section, the only non-specified parameters are 
$\gamma $ and $r_0$, on which the fluctuations strongly depend. 
These will be fitted in this section.
Through the comparison between model and data, 
we have a new method to obtain 
the two point correlation function, which does not require
explicitly the two-point angular
correlation function, do not suffer from 
edge problems, allows an estimation of the errors,
and can be used even with incomplete surveys.
Porciani \& Giavalisco (2002) have also used galaxy counts for this purpose,
but instead they derive the two-point angular correlation function
and then insert it in the Limber equation (which assumes an exponential
two point correlation function in all ranges), so it is halfway between our
method and the standard one.

\begin{figure}
\begin{center}
\mbox{\epsfig{file=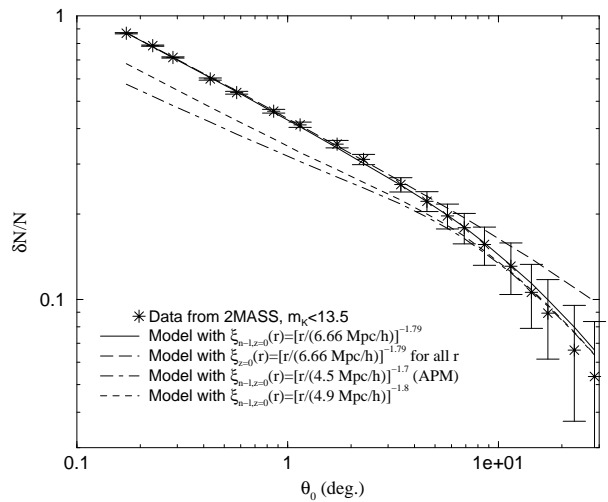,height=6.7cm}}
\end{center}
\caption{Log-log plot: fluctuations of the mean density 
of counts as a function of the radius of the circular regions. 
The error bar of the models is less than 1\% .}
\label{Fig:fluct_area}
\end{figure}

\begin{figure}
\begin{center}
\mbox{\epsfig{file=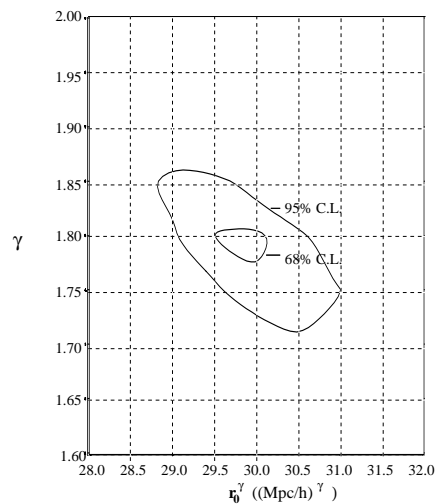,height=7cm,angle=-90}}
\end{center}
\caption{Modified $\chi ^2$-test. The contours show the values of $\gamma $ and
$r_0^\gamma $ which are compatible with the data at 68\% and 95\% C.L.}
\label{Fig:chi2}
\end{figure}

\begin{figure}
\begin{center}
\mbox{\epsfig{file=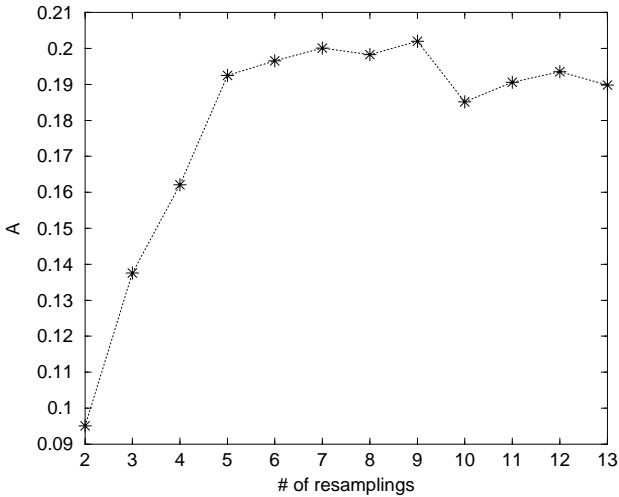,height=6.7cm}}
\end{center}
\caption{Value of $A(C_{ij})$ as a function of the number
of resamplings.}
\label{Fig:a_samp}
\end{figure}

\begin{table*}
\caption{$\delta N/N$ as a function of $\theta _0$ for the 2MASS-data
and the models (same as Fig. \protect{\ref{Fig:fluct_area}}).}
\label{Tab:fluct_area}
\begin{center}
\begin{tabular}{rrrrrr}
$\theta _0$ (deg.) &  2MASS-data  & $r_0=6.66$, $\gamma=1.79$       
& $r_0=6.66$, $\gamma=1.79$ $\forall r$  & $r_0=4.5$, $\gamma=1.7$ &
$r_0=4.9$, $\gamma=1.8$   \\
\hline
0.172 & 0.869$\pm $0.004 & 0.870 & 0.873 & 0.574 &  0.678 \\
0.229 & 0.785$\pm $ 0.004 & 0.777 & 0.780 & 0.521 & 0.607 \\
0.286 & 0.714$\pm $ 0.004 & 0.712 & 0.715 & 0.483 & 0.557 \\
0.43 & 0.599$\pm $ 0.005 & 0.606 & 0.609 & 0.422 & 0.477 \\
0.573 & 0.536$\pm $ 0.006 & 0.540 & 0.544 & 0.383 & 0.427 \\
0.859 & 0.461$\pm $ 0.007 & 0.457 & 0.462 & 0.336 & 0.368 \\
1.146 & 0.413$\pm $ 0.009 & 0.406 & 0.411 & 0.306 & 0.330 \\
1.719 & 0.353$\pm $ 0.011 & 0.342 & 0.348 & 0.270 & 0.287 \\
2.292 & 0.312$\pm $ 0.013 & 0.302 & 0.309 & 0.247 & 0.259 \\
3.44 & 0.254$\pm $ 0.015 & 0.253 & 0.261 & 0.217 & 0.224 \\
4.58 & 0.222$\pm $ 0.018 & 0.221 & 0.231 & 0.195 & 0.200 \\
5.73 & 0.197$\pm $ 0.020 & 0.198 & 0.210 & 0.178 & 0.182 \\
6.88 & 0.179$\pm $ 0.022 & 0.180 & 0.194 & 0.164 & 0.167 \\
8.59 & 0.156$\pm $ 0.024 & 0.158 & 0.176 & 0.146 & 0.148 \\
11.46 & 0.131$\pm $ 0.027 & 0.132 & 0.154 & 0.124 & 0.125 \\
14.32 & 0.106$\pm $ 0.027 & 0.114 & 0.139 & 0.107 & 0.108 \\
17.19 & 0.0895$\pm $ 0.028 & 0.0997 & 0.127 & 0.0946 & 0.0955 \\
22.9 & 0.0662$\pm $ 0.029 & 0.0798 & 0.110 & 0.0762 & 0.0767 \\
28.6 & 0.0535$\pm $ 0.030 & 0.0663 & 0.0984 & 0.0636 & 0.0640 \\
\end{tabular}
\end{center}
\end{table*}

Fig. \ref{Fig:fluct_area} shows the data and some model predictions.
The numbers are in table \ref{Tab:fluct_area}.
The best fit is for $r_0=6.66$ $h^{-1}$Mpc, $\gamma=1.79$. 
Lower values of $r_0$ for the same $\gamma $, such as
$r_0=4.9$ $h^{-1}$Mpc, give considerably less structure than observed
for the present sample of galaxies.
For example,
the two-point correlation function derived from APM optical survey (equal to
that for the mentioned halos):
$\gamma=1.7$, $r_0=4.5$ $h^{-1}$Mpc (Baugh 1996), shows a meaningful defect
of fluctuations.
This leads us to agree with Carlberg et al. (1997),
that optically (blue) selected surveys appear to be significantly less
correlated than K-selected galaxies.

The values of the parameters $\gamma $ and $r_0$ which are compatible
with the data are shown in Fig. \ref{Fig:chi2}, derived from a 
modified $\chi ^2$-test for correlated data (Rubi\~no-Mart\'\i n \&
Betancort-Rijo 2003, secc. 4) applied to the data for $\theta _0<7^\circ$
(the other points have very large error bars and are not useful for
constraining of the parameters; moreover, they are more dependent on the
values of $\xi $ in the linear regime).

In the simplest version of this test ($P_i=0 \ \forall i$, see 
Rubi\~no-Mart\'\i n \& Betancort-Rijo 2003), which for the present
type of problem cannot be far from the best one (determined by
an optimal set of $P_i$), the ordinary uncorrelated $\chi ^2$,
$\chi _{ord}$, and the number of degrees of freedom are rescaled by
a certain factor, $A$, which is a function of the correlations:
\begin{equation}
\chi _{mod}^2=A(C_{ij})\chi _{ord}^2
,\end{equation} 
effective degrees of freedom 
\begin{equation}
n_{mod}=A(C_{ij})\ n_{ord}
,\end{equation} 
where $A$ depends on 
the correlations among data for all pairs of angular distances
[see eq. (22) of Rubi\~no-Mart\'\i n \&
Betancort-Rijo 2003]. To obtain the confidence levels we proceed with
$\chi _{mod}^2$, $n_{mod}$ like in an ordinary uncorrelated test with
$\chi _{ord}^2$, $n_{ord}$
The correlations among data, $C_{ij}=\langle \frac{\delta N}{N}(\theta _i)
\frac{\delta N}{N}(\theta _j)\rangle $ are calculated using resampling
[jackknife resampling (Scranton et al. 2002, Maller et al. 2003); 13
resampling are generated and $\frac{\delta N}{N}$ calculated in each one].
We find $A(C_{ij})=0.190$ (in Fig. \ref{Fig:a_samp} we see how
this value changes very little with the number of resamplings). 

The inferred values of the two parameters are:

\begin{equation}
\gamma=1.79\pm 0.02 \ {\rm (68\% \ C. L.)}\ [\pm 0.07 \ {\rm (95\% \ C. L.)]}
,\end{equation}
\begin{equation}
r_0^\gamma=29.8\pm 0.3 \ {\rm (68\% \ C. L.)}\ [\pm 1.1 \ {\rm (95\% \ C. L.)]}
.\end{equation}
For $\gamma=1.79$, we derive $r_0(z=0)=6.66\pm 0.04$ (68\% C. L.) $h^{-1}$Mpc
(which is represented in Fig. \ref{Fig:fluct_area}).

These values are in excellent agreement 
with the estimations of $\gamma $ in near-infrared
surveys by other means: the angular correlation function is at small scales
proportional to $\theta ^{-0.8}$ in K-band 
(Baugh et al. 1996, Carlberg et al. 1997), which through Limber's equation
gives $\gamma=1.8$. K\"ummel \& Wagner (2000) give an angular correlation
in K-band proportional to $\theta ^{-0.98\pm 0.15}$ which would
give through Limber's equation $\gamma=1.98\pm 0.15$, again consistent
with our value. Maller et al. (2003) with 2MASS K-band data derives 
an angular correlation proportional to $\theta ^{-0.76\pm 0.04}$
which gives $\gamma=1.76\pm 0.04$, very close to our preferred value.
The value of $\gamma $ is also in agreement
with some simulations of $\Lambda CDM$ 
(Primack 2001, Fig. 1: $\gamma =1.6-1.7$)
and the exponents in optical surveys such as APM ($\gamma=1.7$, Baugh 1996).

We get an amplitude $r_0^\gamma(z=0)=29.8\pm 0.3$, more or less in agreement
with the R-selected Las Campanas Redshift Survey which finds 
$r_0^\gamma(z=0.1)=20.5$ (Jing et al. 1998),
equivalent [through eq. (\ref{xiev}) with 
$r_{\rm comoving}=r_{\rm phys.}(1+z)$] to $r_0^\gamma(z=0)=27.0$.
This is also in agreement with the extrapolation
down to $z=0$ of the evolution of $r_0^\gamma $
in K-band selected galaxies at high redshift 
(see Fig. 8 of Carlberg et al. 1997).
However, the blue bands selected surveys give a lower amplitude
of the correlations by a factor of two 
(Carlberg et al. 1997, \S 6 and references therein). 
At high redshifts ($z\approx 3$), 
this difference of amplitude between K-band selected
galaxies and blue-optical selected galaxies is even larger: a factor
3-4 (Daddi et al. 2003). This is consistent because
the correlation length $r_0$ is not independent of the range of
luminosities of the selected sample (Colombo \& Bonometto
2001). Locally it follows roughly a dependence
$r_0\approx 2.2$ $h^{-1}$Mpc $+0.4D_L$ with the scale 
$D_L=[<n>\Phi (>L)]^{-1/3}$ (Colombo \& Bonometto 2001).

Another important remark of Fig. \ref{Fig:fluct_area} is that
the fluctuations of galaxy counts may be accurately predicted by
an expression (\ref{xiz0}) for the correlation. It must be noted
that our method uses a frequentist statistical analysis that not
merely determines the best correlation function but also tells us
how good the best one is. The fact that the best values (of $r_0$,
$\gamma $) are within the 68\% C.L. region tells us that expression
(\ref{xiz0}) is as good as it can possibly be.

No additional effect to large scale structure, for instance intergalactic
extinction, is necessary to explain the variances.
When Zwicky (1957, ch. 3)
observed that $\sigma _N(\theta _0)/\sigma _{Poisson}$ increased with
$\theta _0$ even for large angles, he interpreted this result as
a proof of the presence of an important intergalactic extinction. 
However, as we see in Fig. \ref{Fig:fluct_area}, 
the fact that $\delta N/N$ decreases slowly with increasing $\theta _0$ 
is due to the large scale structure itself, that produces extra fluctuations 
in large scales over those expected from a Poissonian distribution
for scales of several degrees or tens of degrees.

Summing up, we have tested that the $\Lambda $CDM model
of large scale structure is consistent with the observed
galaxy counts variance. This consistency is obtained for
some particular values of the parameters of the 
two point correlation function, so a new method to obtain it indirectly 
without making use of the two-point angular correlation function.
In the application of this method to 2MASS K-band galaxies with $m_K<13.5$
we get $\xi (z=0, r<10\ h^{-1}{\rm Mpc})=(29.8\pm 0.3)r^{-1.79\pm 0.02}$ 
(68\% C.L.).

\

{\bf Acknowledgments:}
Thanks are given to the referee, A. Maller, 
for rather helpful comments and suggestions. Thanks are also given
to N. Sambhus for proof-reading help.
This publication makes use of data products from: 2MASS, which is
a joint project of the Univ. of Massachusetts and the Infrared Processing
and Analysis Center (IPAC), funded by the NASA and the NSF.
This research has made use of the NASA/IPAC Extragalactic
Database (NED) which is operated by the Jet Propulsion Laboratory,
California Institute of Technology, under contract with NASA.
We thank the Swiss National Science Foundation for support under
grant 20-64856.01 .

\end{document}